\documentclass[twoside,12pt]{article}
\usepackage{amsmath,graphicx}

\newcommand{\cd}{\makebox[0.08cm]{$\cdot$}}

\begin{document}

\begin{center}
{\LARGE\textbf{ Nonperturbative renormalization \\ \vspace{0.2cm} in a scalar model within Light-Front Dynamics}}
\end{center}
\vspace{0.3cm}
\begin{center}
\begin{large}
D. Bernard${}^a$,
Th. Cousin${}^b$\footnote{e-mail: cousin@in2p3.fr},
V.A. Karmanov${}^c$\footnote{e-mail: karmanov@sci.lebedev.ru},
J.-F. Mathiot${}^b$\footnote{e-mail: mathiot@in2p3.fr}\\
\end{large}
\vspace{1cm}
{\it
${}^a$ CEA, Centre d'\'etude de Cadarache, 13108 Saint Paul lez Durance Cedex, France\\ \vspace{0.4cm}
${}^b$ Laboratoire de Physique Corpusculaire, Universit\'e Blaise-Pascal,\\ CNRS/IN2P3, 24 avenue des Landais, F-63177 Aubi\`ere Cedex, France\\ \vspace{0.4cm}
${}^c$ Lebedev Physical Institute, Leninsky Prospekt 53, 119991 Moscow, Russia }
\end{center}

\vspace{0.5cm}
\begin{abstract}
Within the covariant formulation of Light-Front Dynamics, in a scalar model with the interaction Hamiltonian
$H=-g\psi^{2}(x)\varphi(x)$, we calculate nonperturbatively the renormalized state vector of a scalar "nucleon"
in a truncated Fock space containing the $N$, $N\pi$ and $N\pi\pi$ sectors. The model gives a simple example
of non-perturbative renormalization which is carried out numerically. Though the mass renormalization
$\delta m^2$ diverges logarithmically with the cutoff $L$, the Fock components of the "physical" nucleon are
stable when $L\to\infty$. \\
\end{abstract}

\vspace{3.5cm}
\noindent  PCCF RI 01-07
\newpage


\section{Introduction}
The knowledge of the hadron properties within the framework of Quantum Chromo-Dynamics (QCD) is one of
the main issue in strong interaction physics.  Several approaches have been pursued in the last twenty years,
in particular lattice gauge calculations.  Among the alternatives to these calculations, Light-Front Dynamics (LFD)
is of particular interest \cite{bpp}.  It has proven successful in many phenomenological applications involving few-body
systems in  particle and nuclear physics. However, the application of LFD to field theoretical calculations is still in its
infancy \cite{npb}.  The main issue to be solved is the renormalization procedure \cite{perr91}. In perturbative
calculations, the renormalization of the electron self-energy in QED, in standard LFD, is already non-trivial in the sense
that it involves non-local counterterms \cite{bro98}.  This unpleasant feature is  however a direct consequence of the
choice of a preferential direction, the $z$ axis, in the determination of the quantization plane.  This can be well
understood in the Covariant Formulation of Light-Front Dynamics (CLFD) \cite{cdkm}, as shown in ref. \cite{kar00}.
In this formulation, the state vector is defined on the light-front surface given by the equation $\omega\cd x=0$,
where $\omega$ is the  four-vector with $\omega^2=0$. The particular case  where $\omega=(1,0,0,-1)$ corresponds
to standard LFD. In CLFD, the counterterm needed to renormalize the electron self energy in first order perturbation
expansion is simply dependent on the orientation of the light front,  defined by the four-vector $\omega$.

We shall investigate in this article how the question of non-perturbative renormalization can be formulated in CLFD.
We shall first derive the general eigenvalue equation, whose solutions are the Fock state components.  As a
first example, we shall illustrate our strategy with a simple model involving two coupled scalar particles.  Namely,
the scalar "nucleon" $N$ radiates the scalar "pions" $\pi$.  In this simple example, the Fock space is restricted to $N$,
$N\pi$ and $N\pi\pi$ states.  Represented as a series of graphs in perturbation theory, it contains an infinite number
of irreducible contributions to the self energy.  They diverge and require renormalization.  At large value of the coupling
constant this system cannot be solved perturbatively.  We show how to determine, in a self-consistent manner, the
non-perturbative mass counterterm.  This counterterm is then calculated numerically.  Models involving two and three
constituent bound states were also analyzed in \cite{ji}.

In a more general field-theoretical framework, in the Light-Front Tamm-Dancoff approximation involving spin $1/2$
particles for instance, renormalization is not reduced to the introduction of a mass counter term. In this case,
one should introduce sector-dependent counter terms, as shown in refs. \cite{perr90}. In the two-nucleon sector,
for spin $1/2$ particles, additional box divergences appear \cite{gl-wils}. These divergences and the sector
dependent counter terms are absent in the scalar model restricted to  the dressed "one-scalar-nucleon" system.
Therefore, their analysis is beyond the scope of the present paper. Inspite of that, the scalar model is still rather
instructive, since the renormalization considered here is not reduced to the perturbative one. We will consider very
large coupling constant, excluding the convergence of the perturbative series.

The plan of the article is as follows.  In section \ref{eigen} we establish the general equations of motion for the Fock
components.  In the truncated Fock space,  the corresponding system of equations, which determine the Fock
components and the mass renormalization, is detailed in sect. \ref{equat}. The renormalization of the wave function
and of the coupling constant  is calculated in sections \ref{renwf} and \ref{rcc} respectively.  Numerical results are
presented and discussed in section \ref{num}.   We present our conclusions in section \ref{concl}.


\section{Eigenstate equation}\label{eigen}
We start with the general eigenstate equation for the state vector \cite{cdkm}:
	\begin{equation}\label{eq1}
	\hat{P}^2\ \phi(p)=M^2\ \phi(p),
	\end{equation}
where
	\begin{equation}\label{kt1}
	\hat{P}_{\mu}= \hat{P}^0_{\mu} +\hat{P}^{int}_{\mu}\ .
	\end{equation}
We have decomposed here the momentum operator $\hat{P}_{\mu}$ into two parts: the free one, $\hat{P}^0_{\mu}$,
and the interacting one $\hat{P}^{int}_{\mu}$, given by:
	\begin{eqnarray}\label{kt3}
	\hat{P}^0_{\mu} &=&\sum_{i}\int d^\dagger_{i} (\vec{k})d_{i}(\vec{k})k_{\mu}\ d^3k\ , \nonumber\\
	\hat{P}^{int}_{\mu}&=&\omega_{\mu}\int H^{int}(x)\delta(\omega\cd x) \ d^4x=\omega_{\mu}\int_{-\infty}^{+\infty} \tilde{H}^{int}(\omega\tau)\frac{d\tau}{2\pi} \ ,
	\end{eqnarray}
where we have denoted by $\tilde{H}^{int}$ the Fourier transform of  the interaction Hamiltonian:
	\begin{equation}\label{rul4}
	\tilde{H}^{int}(p)=\int H^{int}(x)\exp(-ip\cd x)d^4x\ ,
	\end{equation}
and $d^\dagger_{i}$ ($d_{i}$) corresponds to the creation (destruction) operator for the various particles under
consideration. The explicitly covariant formulation  of LFD manifests itself in the fact that $\hat{P}^{int}_{\mu}$ in
eq.  (\ref{kt3}) is proportional to $\omega_{\mu}$ and is determined by the integral over the light-front plane
$\omega\cd x=0$.

The equations for the Fock components can be obtained from (\ref{eq1}), by substituting there the Fock decomposition
for the state vector  $\phi(p)$ and calculating the matrix elements of $\hat{P}^2$ in the Fock  space.  With the above
expressions for $\hat{P}$, eq. (\ref{eq1}) obtains the form:
	\begin{equation}\label{eq1a}
	\left[(\hat{P}^{0})^2+(\omega\cd\hat{P}^{0}) \int \tilde{H}^{int}(\omega\tau)\frac{d\tau}{2\pi} +\int \tilde{H}^{int}(\omega\tau)\frac{d\tau}{2\pi}  (\omega\cd\hat{P}^{0} )\right] \phi(p)=M^2\ \phi(p)\ .
	\end{equation}
In order to simplify this equation, we can use the fact that the operators $(\omega\cd\hat{P}^{0} )$ and
$\int\tilde{H}^{int}(\omega\tau)d\tau$ commute. Indeed, from the commutation relation $[\hat{P}_{\mu},\hat{P}_{\nu}]=0$
we get:
	$$ [\omega\cd\hat{P},\hat{P}_{\nu}]= [\omega\cd\hat{P}^{0},\hat{P}^{0}_{\nu}+\hat{P}^{int}_{\nu}]= [\omega\cd\hat{P}^{0},\hat{P}^{int}_{\nu}]=0\ . $$
Moreover, since $\omega^2=0$, we can replace here and below $(\omega\cd\hat{P}^{0})$ by $(\omega\cd\hat{P})$.
We thus obtain:
	$$ (\omega\cd\hat{P}^{0}) \int \tilde{H}^{int}(\omega\tau) \phi(p) d\tau =\int (\omega\cd\hat{P}) \tilde{H}^{int}(\omega\tau) \phi(p) d\tau =(\omega\cd p)\int \tilde{H}^{int}(\omega\tau)\phi(p) d\tau \ , $$
and the equation (\ref{eq1a}) is transformed to:
	\begin{equation}\label{eq1b}
	2(\omega\cd p)\int \tilde{H}^{int}(\omega\tau)\frac{d\tau}{2\pi} \phi(p)= -\left[(\hat{P}^{0})^2-M^{2}\right]\phi(p)\ .
	\end{equation}
The state vector  $\phi(p)$ is now decomposed in Fock components according to:
	\begin{eqnarray}\label{wfp1}
	\phi(p) &=
	&(2\pi)^{3/2}\int\phi_{1}(k_1,p,\omega\tau) a^\dagger(\vec{k}_1)|0\rangle \delta^{(4)}(k_1-p-\omega\tau) 2(\omega\cd p)d\tau \frac{d^3k_1}{(2\pi)^{3/2}\sqrt{2\varepsilon_{k_1}}} \nonumber \\
	&+&(2\pi)^{3/2}\int\phi_{2}(k_1,k_2,p,\omega\tau) a^\dagger(\vec{k}_1)b^\dagger(\vec{k}_2)|0\rangle \nonumber \\
	&\times& \delta^{(4)}(k_1+k_2-p-\omega\tau) 2(\omega\cd p)d\tau \frac{d^3k_1}{(2\pi)^{3/2}\sqrt{2\varepsilon_{k_1}}} \frac{d^3k_2}{(2\pi)^{3/2}\sqrt{2\varepsilon_{k_2}}} \nonumber \\
	&+&(2\pi)^{3/2}\int{\phi_{3}}(k_1,k_2,k_3,p,\omega\tau) a^\dagger(\vec{k}_1)b^\dagger(\vec{k}_2) b^\dagger(\vec{k}_3)|0\rangle \nonumber \\
	&\times& \delta^{(4)}(k_1+k_2+k_2-p-\omega\tau) 2(\omega\cd p)d\tau \nonumber \\
	&\times& \frac{d^3k_1}{(2\pi)^{3/2}\sqrt{2\varepsilon_{k_1}}} \frac{d^3k_2}{(2\pi)^{3/2}\sqrt{2\varepsilon_{k_2}}} \frac{d^3k_3}{(2\pi)^{3/2}\sqrt{2\varepsilon_{k_3}}} + \cdots\ ,
	\end{eqnarray}
where $\varepsilon_{k_{i}}=\sqrt{\vec{k_{i}}^{2}+m_{i}^{2}}$ and $m_{i}$ is the mass of the particle $i$ of momentum
$k_{i}$. We introduce in (\ref{wfp1}) spinless particles of two types: the "nucleon" (with creation operator $a^\dagger$)
and the "pion" (with creation operator $b^\dagger$).  The state vector (\ref{wfp1}) represents the "dressed nucleon",
consisting of the "bare nucleon" and the admixture of one, two,\ldots, many "pions".  Our dressed nucleon is also
scalar, therefore the state vector  (\ref{eq1b}) corresponds to zero total angular momentum. This means  that the Fock
components are scalars and depend only on scalar  products of all available four-vectors.

The conservation law for the momenta in each Fock component has the form:
	$$ k_1+k_2+\cdots +k_{n}=p+\omega\tau\ . $$
Hence, the action of the operator $(\hat{P}^{0})^2-M^2$ in (\ref{eq1b}) on the state vector $\phi(p)$ is reduced
to the multiplication of each Fock component by the factor $(\sum k_i)^2-M^{2}=2(\omega\cd p)\tau$.

If we introduce the notation:
	$$ {\cal G}(p)=2(\omega\cd p)\hat{\tau}\phi(p)\ , $$
where $\hat \tau $ is the operator which, acting on a given component $\phi_i$ of $\phi$, gives $\tau \phi_{i}$.
${\cal G}$ has therefore a Fock decomposition which differs from (\ref{wfp1}) by the replacement of the wave
functions $\phi_{i}$ by the vertex parts $\Gamma_{i}$ given by:
	\begin{equation}\label{eq2}
	\Gamma_{i}\equiv 2(\omega\cd p)\tau\phi_{i}=(s-M^{2})\phi_{i}\ ,
	\end{equation}
where $s=(\sum k_i)^2$. We thus find the eigenvalue equation:
	\begin{equation}\label{eq3}
	\frac{1}{2\pi}\int \tilde{H}^{int}(\omega\tau)\frac{d\tau}{\tau} {\cal G}(p)= -{\cal G}(p)\equiv -\lambda(M^2){\cal G}(p)\ .
	\end{equation}
We introduce in (\ref{eq3}) the factor $\lambda(M^2)$ depending on $M^{2}$.  The eigenvalue $M^{2}$ is found
from the condition $\lambda(M^2)=1$.  This equation is quite general and equivalent to the eigenstate equation
(\ref{eq1}).


\section{Equation for the Fock components}\label{equat}

\subsection{System of coupled integral equation}
For the simplified model we consider in this study, we take the following interaction Hamiltonian:
	\begin{equation}\label{eq4}
	H=-g\psi^{2}(x)\varphi(x)\,
	\end{equation}
where the scalar field $\psi$ with mass $m$ corresponds to  the scalar "nucleon" and the field $\varphi$ with mass
$\mu$ corresponds to  the scalar "pion".

	\begin{figure}[!htb]
	\centering
	\includegraphics[height=7cm,clip=true]{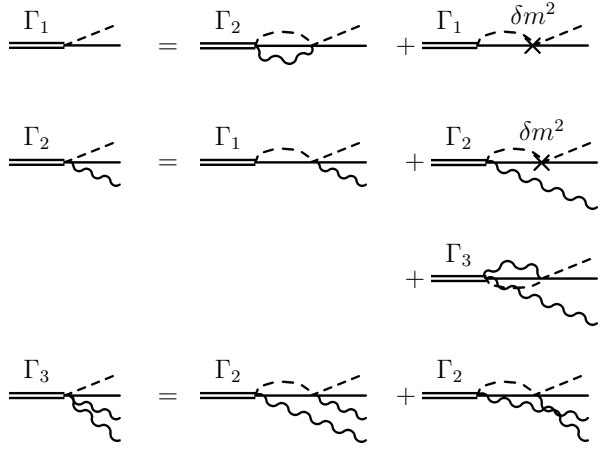}
	\caption{\label{fig1} Diagrammatical representation of the eigenvalue equation projected onto the Fock
	components of the state vector (restricted here to N=3). For the sake of clarity, we remove all  kinematical
	variables.}
	\end{figure}

The system of equations for these vertex parts is shown graphically in fig. \ref{fig1}.  In order to write down this
system of equations, it is enough to apply to the diagrams of fig.  \ref{fig1} the rules of the graph technique detailed
in ref.\cite{cdkm}.  We thus find:
	\begin{subequations} \label{eq5_3_1}
	\begin{eqnarray}
	\Gamma_1 (k_{1},p,\omega\tau_1)
	&=&\delta m^{2} \int \Gamma_{1}(k_1',p,\omega\tau')\delta^{(4)}(k_1'-p-\omega\tau') \theta(\omega\cd k'_{1})\delta({k'}_{1}^{2}-m^2)d^{4}k'_{1} \frac{d\tau'}{\tau'-i0} \nonumber\\
	&+& g \frac{1}{(2\pi)^{3}}\int\Gamma_{2}(k'_1,k'_2,p,\omega\tau') \delta^{(4)}(k'_{1}+k'_{2}-p-\omega\tau') \nonumber\\
	&\times& \theta (\omega\cd k'_{1})\delta({k'}_{1}^{2}-m^2)d^{4}k'_{1} \theta (\omega\cd k'_{2})\delta({{k'}_{2}}^{2}-\mu^2)d^{4}k'_{2} \frac{d\tau'}{\tau'-i0}\ ,\label{eq5_3_1.a} \\
	&& \nonumber \\
	&& \nonumber \\
	\Gamma_{2}(k_1,k_2,p,\omega\tau_2)
	&=& g\int \Gamma_{1}(k'_1,p,\omega\tau')\delta^{(4)}(k'_1-p-\omega\tau') \theta(\omega\cd k'_{1})\delta({k'}_{1}^{2}-m^2)d^{4}k'_{1} \frac{d\tau'}{\tau'-i0} \nonumber\\
	&+& \delta m^{2} \int \Gamma_{2}(k'_1,k_2,p,\omega\tau') \delta^{(4)}(k_1'+k_2-p-\omega\tau') \nonumber\\
	&\times& \theta(\omega\cd k_{1}')\delta({k'}_{1}^{2}-m^2)d^{4}k_{1}' \frac{d\tau'}{\tau'-i0} \nonumber\\
	&+& g \frac{1}{(2\pi)^{3}}\int \Gamma_{3}(k_1',k_2,k_3',p,\omega\tau') \delta^{(4)}(k_{1}'+k_2+k_{3}'-p-\omega\tau') \nonumber\\
	&\times& \theta (\omega\cd k_{1}')\delta({k'}_{1}^{2}-m^2)d^{4}k_{1}' \theta (\omega\cd k_{3}')\delta({k'}_{3}^{2}-\mu^2)d^{4}k_{3}' \frac{d\tau'}{\tau'-i0}\ ,\label{eq5_3_1.b}\\
	&&\nonumber\\ && \nonumber \\
	\Gamma_{3}(k_1,k_2,k_3,p,\omega\tau_3)
	&=& g\int \Gamma_{2}(k_1',k_2,p,\omega\tau') \delta^{(4)}(k_1'+k_2-p-\omega\tau') \nonumber\\
	&\times& \theta(\omega\cd k_{1}')\delta({k'}_{1}^{2}-m^2)d^{4}k_{1}' \frac{d\tau'}{\tau'-i0} \nonumber\\
	&+& g\int \Gamma_{2}(k_1',k_3,p,\omega\tau') \delta^{(4)}(k_1'+k_3-p-\omega\tau') \nonumber\\
	&\times& \theta(\omega\cd k_{1}')\delta({k'}_{1}^{2}-m^2)d^{4}k_{1}' \frac{d\tau'}{\tau'-i0}\ .\label{eq5_3_1.c}
	\end{eqnarray}
	\end{subequations}

\begin{figure}
	\centering
	\includegraphics[height=4.5cm,clip=true]{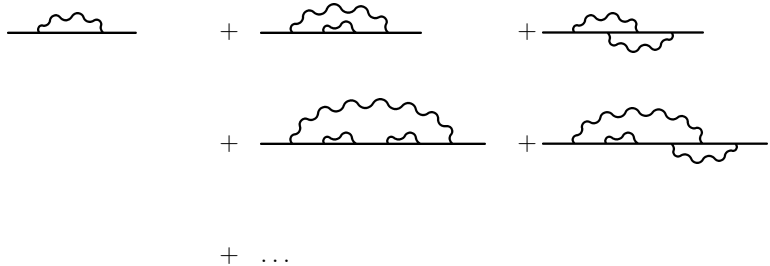}
	\caption{\label{fig2} Perturbative expansion, in terms of the pion-nucleon coupling constant $g$, of the nucleon self-energy.}
	\end{figure}

\begin{figure}
	\centering
	\includegraphics[height=2cm,clip=true]{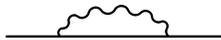}
	\caption{\label{fig3} First order perturbative expansion of the nucleon self-energy.}
	\end{figure}

The origin of the mass counterterm $\delta m^2$ in these equations is explained below.  Since we truncate the Fock
space to three particles, we omit in the last equation (\ref{eq5_3_1.c}) the coupling of $\Gamma_{3}$ to the four-body
component $\Gamma_{4}$.  Note that our approximation is not based on a perturbative expansion in terms of the
coupling constant.  It is based on a decomposition over intermediate states with an increasing number of particles.
Once this number is fixed, we solve a non-perturbative problem in terms of $g$.  If we iterate the system (\ref{eq5_3_1}),
i.e., express $\Gamma_{3}$ and $\Gamma_{2}$ from the third and the second equations and substitute it in the first one,
we get for $\Gamma_{1}$ the sum of the diagrams shown on fig.  \ref{fig2}.  Of course, this series of graphs corresponds
to all the irreducible contributions to the self energy  with intermediate states up to $N\pi\pi$.  These contributions are
iterated again, i.e., they appear repetitively on the nucleon line.  In contrast to the case of the intermediate state $N\pi$,
which generates only one self energy diagram shown on fig.  \ref{fig3}, the number of irreducible contributions generated
by the intermediate states up to $N\pi\pi$ is infinite.  The system of equations (\ref{eq5_3_1}) corresponds to the sum of
all of them.

In the right-hand side of equations (\ref{eq5_3_1.a}) and (\ref{eq5_3_1.b}) we introduced the counterterm $\delta m^{2}$
corresponding to the interaction Hamiltonian $H=-\delta m^2 \psi^2(x)$.  This term provides mass renormalization.
The counterterm is however not introduced in the third equation containing the three-body intermediate state.
At this point we have some freedom in the definition of the model Hamiltonian.  In principle, the counterterm could be
also introduced into the three-body intermediate state, since formally it does not increase the number of particles in the
intermediate state.  Our motivation in choosing the Hamiltonian without the counterterm in the last $n$-th Fock sector
is the following.  Consider, for example, the mass operator $-g^2\Sigma(p^2)$ in second order perturbation theory.
It is given by the diagram of fig. \ref{fig3} with two particles ($N$ and $\pi$) in the intermediate state. The counterterm
$\delta m^2$, which is shown by a cross on the nucleon line (one particle $N$ in the intermediate state) just renormalizes
this mass operator.  So, one "pion" is deleted when the counterterm appears.  Therefore, for the renormalization of the
mass operator determined by the sum of (an infinite number of) irreducible diagram with $n$ particles in the intermediate
state, one should consider the graphs, with the counterterm insertions, with $n-1$ particles in the intermediate states
only \cite{bro98}.  These diagrams are just generated by the Hamiltonian without the counterterm in the $n$-th Fock sector.

We can easily transform eqs. (\ref{eq5_3_1}) by performing the integrations which do not involve loops, keeping the loop
integrals untouched.  The result is the following:
	\begin{subequations}\label{eq5_3}
	\begin{eqnarray}\label{eq5_3.a}
	\Gamma_1 (k_{1},p,\omega\tau_1)
	&=& \frac{\delta m^{2}}{2(\omega\cd p)\tau_1}\Gamma_{1}(k_1,p,\omega\tau_1) \nonumber\\
	&+& g \frac{1}{(2\pi)^{3}}\int \Gamma_{2}(k_1',k_2',p,\omega\tau') \delta^{(4)}(k_{1}'+k_{2}'-p-\omega\tau') \nonumber\\
	&\times& \theta (\omega\cd k_{1}')\delta({k'}_{1}^{2}-m^2)d^{4}k_{1}' \theta (\omega\cd k_{2}')\delta({k_{2}'}^{2}-\mu^2)d^{4}k_{2}' \frac{d\tau'}{\tau'-i0}\ ,
	\end{eqnarray}
	\begin{eqnarray}\label{eq5_3.b}
	\Gamma_{2}(k_1,k_2,p,\omega\tau_2)
	&=& \frac{g}{2(\omega\cd p)\tau_1}\Gamma_{1}(p_1,p,\omega\tau_1) \nonumber\\
	&+& \frac{\delta m^{2}}{2(\omega\cd p )\tau_2 x_1} \Gamma_{2}(k_1,k_2,p,\omega\tau_2) \nonumber\\
	&+& g \frac{1}{(2\pi)^{3}}\int \Gamma_{3}(k_1',k_2,k_3',p,\omega\tau') \delta^{(4)}(k_{1}'+k_2+k_{3}'-p-\omega\tau') \nonumber\\
	&\times& \theta (\omega\cd k_{1}')\delta({k'}_{1}^{2}-m^2)d^{4}k_{1}' \theta (\omega\cd k_{3}')\delta({k_{3}'}^{2}-\mu^2)d^{4}k_{3}' \frac{d\tau'}{\tau'-i0}\ , \\
	&&\nonumber \\ \label{eq5_3.c} \Gamma_{3}(k_1,k_2,k_3,p,\omega\tau_3)&=& \frac{g}{2(\omega\cd p)\tau_2'(1-x_2)} \Gamma_{2}(k_1',k_2,p,\omega\tau_2') \nonumber\\
	&+& \frac{g}{2(\omega\cd p)\tau_2''(1-x_3)} \Gamma_{2}(k_1'',k_3,p,\omega\tau_2'')\ .
	\end{eqnarray}
	\end{subequations}
In eqs. (\ref{eq5_3.a}) and  (\ref{eq5_3.b}), $2(\omega\cd p)\tau_1=k_1^2-p^2=m^2-p^2\to 0$ when $p^2\to m^2$.
Since $\tau_1$ is in the denominator, we keep $p^2\neq m^2$ and take the limit $p^2\to m^2$ in the final equation.
In eq. (\ref{eq5_3.b}), $p_1=p-\omega \tau_1$ with $2(\omega\cd p)\tau_2=s_{12}-m^2$ and $s_{12}=(k_1+k_2)^2$.
In eq. (\ref{eq5_3.c}) we use the notations:
	$$2(\omega\cd p)\tau_2'=s_{12}'-m^2, \quad 2(\omega\cd p)\tau_2''=s_{12}''-m^2,$$
where
	$$s_{12}'=(k_1'+k_2)^2,\quad s_{12}''=(k_1''+k_3)^2$$
and $k_1',k_1''$ are determined by the conservation laws:
	\begin{equation}\label{conserv}
	k_1'=k_1+k_3+\omega\tau_2'-\omega\tau_{123},\quad k_1''=k_1+k_2+\omega\tau_2''-\omega\tau_{123}\ ,
	\end{equation}
where $\omega \tau_{123}$ is the momentum of the spurion line  entering the diagram. Everywhere we note
$x_{i}=\omega\cd k_{i}/\omega\cd p$ .


\subsection{Reduction to two-particle Fock states}
\begin{figure}
	\centering
	\includegraphics[height=4cm,clip=true]{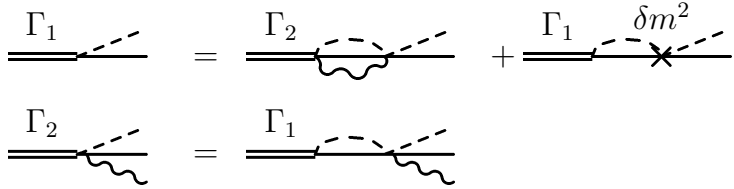}
	\caption{\label{fig4} Diagrammatical representation of the eigenvalue equation in first order perturbation theory. }
	\end{figure}

We consider first the approximation in which the state vector (\ref{wfp1}) contains only the bare nucleon $N$ and the
state $N\pi$.  In this approximation, the system of equations for one-body and two-body Fock components is represented
diagrammatically on fig.  \ref{fig4}.  It is easily obtained by omitting $\Gamma_3$ in (\ref{eq5_3_1.c}), together with the
counterterm in the equation which determines the last Fock sector $\Gamma_{2}$.  This truncation of the Fock space,
retaining the minimal number of components, is equivalent to second order perturbation theory.

Rewritten in terms of the variables $\vec{q}$ and $\vec{n}$ (see appendix A), this system of equations obtains the
simple form:
	\begin{subequations}\label{eq6}
	\begin{eqnarray}
	\Gamma_1 &=&\frac{\delta m_{0}^{2}}{m^{2}-p^{2}}\Gamma_1+ g\int \sigma(\vec{q}\,',p^2)\ \Gamma_{2}(\vec{q}\,',\vec{n})\ \frac{d^{3}q'}{(2\pi)^{3}}\ , \\
	\Gamma_2(\vec{q},\vec{n})&=&\frac{g}{m^{2}-p^{2}}\ \Gamma_1\ ,
	\end{eqnarray}
	\end{subequations}
where $\sigma(\vec{q},p^2)$ is the integrand which determines the self-energy $\Sigma(p^2)$:
	\begin{eqnarray} \Sigma(p^2)&=&\int \sigma(\vec{q'},p^2)\frac{d^{3}q'}{(2\pi)^{3}}\ , \label{eq7a} \\
	\sigma(\vec{q},p^2)&=&\frac{1}{2}\frac{\sqrt{s_{12}}}{(s_{12}-p^{2}-i0)} \frac{1}{\varepsilon(q,m)\varepsilon(q,\mu)}\ , \nonumber
	\end{eqnarray}
with
	\begin{equation}\label{s}
	s_{12}=(k_{1}+k_{2})^{2}=\left[\varepsilon(\vec{q},m)+\varepsilon(\vec{q},\mu) \right]^{2}\ ,
	\varepsilon(\vec{q},m)=\sqrt{m^{2}+\vec{q}^{2}}\ ,
	\end{equation}
and similarly for $\varepsilon(\vec{q},\mu)$. Since $\Gamma_{1}$ does not depend on the relative momentum,
it follows  from eqs. (\ref{eq6}) that $\Gamma_2(\vec{q},\vec{n})$ does not depend on the relative momentum too,
i.e., $\Gamma_2(\vec{q},\vec{n})=const$. Hence,  we get:
	\begin{eqnarray}\label{eq8}
	\left(-1+\frac{\delta m_{0}^{2}}{m^{2}-p^{2}}\right) \Gamma_1 &+& g\Sigma(p^{2})\Gamma_{2}=0\ , \nonumber\\
	\frac{g}{m^{2}-p^{2}}\Gamma_{1}&-&\Gamma_{2}=0\ .
	\end{eqnarray}
From (\ref{eq8}) we find the eigenvalue equation:
	\begin{equation}\label{eq9}
	p^{2}=m^{2}-\delta m_{0}^{2}-g^{2}\Sigma(p^{2})\ .
	\end{equation}
The counter term $\delta m_{0}^{2}$ is determined from the on-shell condition $p^{2}=m^{2}$, where $m$ is the
physical mass of the nucleon.  This gives:
	\begin{equation}\label{eq10}
	\delta m_{0}^{2}=-g^{2}\Sigma(m^{2})\ ,
	\end{equation}
as expected in  second order calculation of  mass renormalization.


\subsection{Solution for the three particle system}

	\begin{figure}
	\centering
	\includegraphics[height=4.5cm,clip=true]{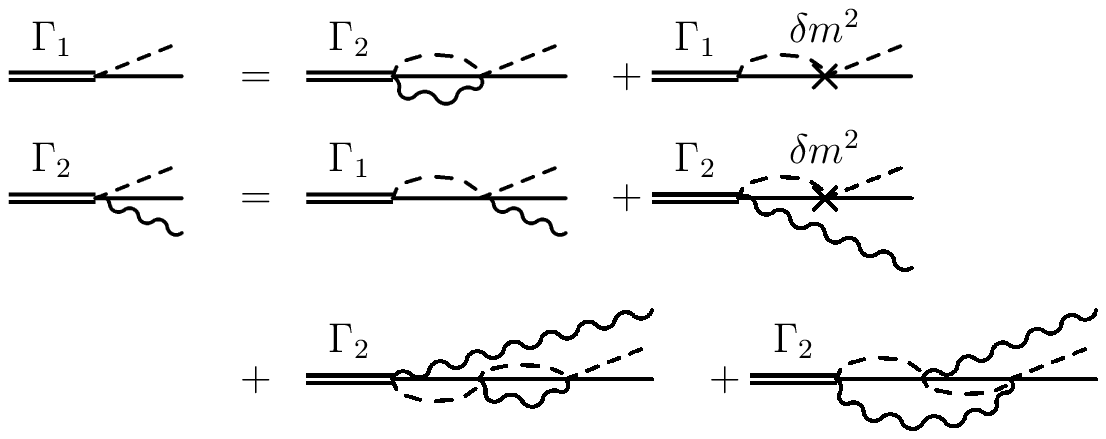}
	\caption{\label{fig5} Same as Figure 1, but where the last Fock sector (n=3) has been expressed in terms of the
	two-body one.}
	\end{figure}

Consider now the system of equations which incorporates the three-body Fock component $\Gamma_{3}$.
It was shown graphically on fig. 1. We can easily express $\Gamma_{3}$ through $\Gamma_{2}$. The system
of equations is thus transformed as shown in fig. 5. Its analytical representation has the form:
	\begin{subequations} \label{eq11}
	\begin{eqnarray}
	\Gamma_1
	&=&\frac{\delta m^{2}}{m^{2}-p^{2}}\Gamma_1+ g\int \sigma(\vec{q}\,',p^2)\Gamma_{2}(\vec{q}\,',\vec{n})\frac{d^{3}q'}{(  2\pi)^{3}}\ , \label{eq11.a}\\
	\Gamma_2(\vec{q},\vec{n})
	&=&\frac{g}{m^{2}-p^{2}}\Gamma_1 + \frac{\delta m^2+g^{2}\Sigma(s_{1})} {(s_{12}-p^{2})x_{k_{1}}}\Gamma_2(\vec{q},\vec{n})  \nonumber \\
	&+&g^{2}\int \Pi(\vec{q},\vec{q}\,',\vec{n},p^{2})\sigma(\vec{q}\,',p^2) \Gamma_{2}(\vec{q}\,',\vec{n})\frac{d^{3}q'}{(2\pi)^{3}}\ . \label{eq11.b}
	\end{eqnarray}
	\end{subequations}
Each term in eqs. (\ref{eq11}) is represented by a graph on fig. 5. In these equations, $s$ is given by (\ref{s}), and:
	$$ s_{1}= (k_{1}-\omega\tau)^{2}=m^{2}-x_{k_{1}}(s_{12}-p^{2})\ , $$
	$$ x_{k_{1}}=\frac{\omega\cd k_{1}}{\omega\cd p}=\frac{1}{\sqrt{s_{12}}} \left[\varepsilon(\vec{q},m)-\vec{n}\cd \vec{q}\right]\ . $$
In this equation, $\Pi(\vec{q},\vec{q}\,',\vec{n},p^{2})$ is the propagator of the second intermediate state:
	\begin{eqnarray}\label{eq12}
	\Pi(\vec{q},\vec{q}\,',\vec{n},p^{2})
	&=& \int\theta\left[\omega\cd (k_{1}-k_{2}')\right] \delta\left[(k_{1}-k_{2}'+\omega\tau_{2}-\omega\tau'')^{2}-m^{2}\right] \frac{d\tau''}{\tau''-i0} \nonumber\\
	&=& \frac{\theta\left[\omega\cd (k_{1}'-k_{2})\right]} {m^{2}-(k_{1}'-k_{2}-\omega\tau')^{2}}\ .
	\end{eqnarray}
Its expression in terms of the variables $\vec{q},\vec{q}\,',\vec{n}$ can be easily calculated with the kinematics detailed
in the appendix.

The system of equation (\ref{eq11}) can now be solved by two independent methods.

{\it i)} The first one consists in the elimination of $\Gamma_{1}$ in the two equations in (\ref{eq11}).  The result is
an equation for $\Gamma_{2}$ which, at $p^{2}=m^{2}$, obtains the form:
	\begin{eqnarray}\label{eq11a}
	\Gamma_2(\vec{q},\vec{n})
	&=& -\frac{g^{2}}{\delta m^{2}} \int \sigma(\vec{q}\,',p^2)\Gamma_{2}(\vec{q}\,',\vec{n})\frac{d^{3}q'}{(  2\pi)^{3}} \nonumber\\
	&+& \frac{g^{2}\Sigma(s_{1})+\delta  m^2}{(s_{12}-p^{2})x_{k_{1}}}\Gamma_2(\vec{q},\vec{n}) +g^{2}\int \Pi(\vec{q},\vec{q}\,',\vec{n},p^{2})\sigma(\vec{q}\,',p^2) \Gamma_{2}(\vec{q}\,',\vec{n})\frac{d^{3}q'}{(2\pi)^{3}} \nonumber\\
	&\equiv&\lambda(\delta m^{2})\Gamma_2(\vec{q},\vec{n})\ .
	\end{eqnarray}
Like in eq. (\ref{eq3}), we introduce in (\ref{eq11a}) the factor $\lambda(\delta m^{2})$. The mass renormalization
$\delta m^{2}$ will be found from the non-perturbative condition $\lambda(\delta m^{2})=1$.  {\it ii)} The second method
consists in the direct elimination of  $\Gamma_{2}$ in the two equations (\ref{eq11}). We can indeed rewrite eq.
(\ref{eq11.b}) in the following form:
	\begin{eqnarray}
	\left[\Gamma_2(\vec{q},\vec{n})-\frac{\delta m^2+g^{2}\Sigma(s_{1})} {(s_{12}-p^{2})x_{k_{1}}}\Gamma_2(\vec{q},\vec{n}) -g^{2}\int \Pi(\vec{q},\vec{q}\,',\vec{n},p^{2})\sigma(\vec{q}\,',p^2)\Gamma_{2}(\vec{q}\,' ,\vec{n})\frac{d^{3}q'}{(2\pi)^{3}}\right]=\nonumber \\
	\frac{g}{m^{2}-p^{2}}\Gamma_1\ .
	\end{eqnarray}
After discretization of the momenta, this equation can be written schematically:
	\begin{equation}
	A_{ij} \Gamma_{2}^j=\frac{g}{m^2-p^2}\Gamma_{1} \left[1\right]^i\ ,
	\end{equation}
where $\left[1\right]$ is the vector whose every component is 1, and $A_{ij}$ is a two-dimensional matrix obtained
after discretization of the momenta (index $i$ for $q, \vec{n} \cd \vec{q}$ and index $j$ for $q', \vec{n}.\vec{q'}$).
The vertex function $\Gamma_{2}$ can now be expressed in terms of  $\Gamma_{1}$ after a simple matrix inversion:
	\begin{equation}
	\Gamma_{2}^i=\frac{g}{m^2-p^2}\Gamma_{1} A_{ij}^{-1}\left[1\right]^j\ .
	\end{equation}
After insertion in eq.(\ref{eq11.a}), we  end up with an  equation involving $\Gamma_{1}$ only. Since $\Gamma_{1}$ is
a  non-zero constant, it can be removed from the equation, leading to an  equation for $\delta m^2$ generalizing
eq. (\ref{eq10}). In the perturbative  limit, the matrix $A$ reduces to the unit matrix, and we therefore  recover
eq. (\ref{eq10}) exactly.


\section{Renormalization of the wave function}\label{renwf}

\subsection{Non-perturbative case}
We schematically rewrite the state vector $|p\rangle$, given by eq. (\ref{wfp1}), as:
	$$ \vert p\rangle =\phi_1\vert N\rangle +\phi_2\vert N\pi\rangle + \phi_3\vert N\pi\pi\rangle\ . $$
It is normalized as follows \cite{cdkm}:
	\begin{equation}\label{nor1}
	\langle p' \vert p\rangle =2p_0\ \delta^{(3)}(\vec{p}- \vec{p}\,')\ .
	\end{equation}
The Fock components are thus normalized in order to provide the condition (\ref{nor1}). Substituting the state
vector (\ref{wfp1}) in the left-hand side of eq. (\ref{nor1}), we get, before normalization of  the Fock components:
	\begin{equation}\label{nor1a}
	\langle p' \vert p\rangle = Z\  2p_0\ \delta^{(3)}(\vec{p}- \vec{p}\,')\ ,
	\end{equation}
where
	$$Z=N_{1}+N_{2}+N_{3},$$
with:
	\begin{subequations} \label{nor2}
	\begin{eqnarray}
	N_1&=&\phi_1^2\ , \label{nor2.a} \\
	N_2&=& \frac{1}{2(2\pi)^3}\int \phi_2^2(\vec{q},\vec{n}) \frac{[\varepsilon(q,m)+\varepsilon(q,\mu)]} {\varepsilon(q,m)\varepsilon(q,\mu)} d^3q\ , \label{nor2.b} \\ N_3&=&\frac{1}{4(2\pi)^6}\int \phi_3^2(\vec{q}_1,\vec{q}_2,\vec{q}_3,\vec{n}) \delta^{(3)}(\vec{q}_1+\vec{q}_2+\vec{q}_3) \nonumber \\
	&\times& \frac{[\varepsilon(q_1,m)+\varepsilon(q_2,\mu)+\varepsilon(q_3,\mu)]} {\varepsilon(q_1,m)\varepsilon(q_2,\mu)\varepsilon(q_3,\mu)} d^3q_1 d^3q_2 d^3q_3 \ .\label{nor2.c}
	\end{eqnarray}
	\end{subequations}
From $\Gamma_2$ calculated with eq. (\ref{eq11a}), we can find $\phi_2$ according to:
	\begin{equation}\label{psi2}
	\phi_2(\vec {q},\vec{n})= \frac{\Gamma_2(\vec {q},\vec{n})}{s_{12}-m^2}\ .
	\end{equation}
The calculation of $N_2$ is then straightforward. To calculate $N_{1}$ and $N_{3}$  we should know $\phi_{1}$
and $\phi_{3}$. The calculation of $\phi_{1}$ and $\phi_{3}$ is explained in appendix \ref{psi13}. Note that $\phi_1$
does not depend on any relative momentum.  The normalized state vector satisfying the condition (\ref{nor1}) obtains
the form:
	$$ \vert p\rangle =\phi^{ren}_1\vert N\rangle +\phi_2^{ren}\vert N\pi\rangle + \phi_3^{ren}\vert N\pi\pi\rangle\ , $$
where
	\begin{equation}\label{nor1b}
	\phi^{ren}_{1,2,3}=\phi_{1,2,3}/\sqrt{Z}\ .
	\end{equation}
We can introduce the creation  operator of the new, composed field, creating directly the state $|p\rangle$. It is
written schematically as:
	\begin{equation}\label{nor1c}
	A^\dagger(\vec{p}) =\phi^{ren}_1 a^\dagger +\phi_2^{ren} a^\dagger b^\dagger + \phi_3^{ren} a^\dagger b^\dagger b^\dagger \ ,
	\end{equation}
so that $|p\rangle=A^\dagger(\vec{p})|0>$. With the renormalized wave functions $\phi^{ren}$, the vacuum expectation
value of the  commutator: $\langle 0|\left[A(\vec{p}),A^\dagger(\vec{p})\right]|0  \rangle =2p_0 \delta^{(3)}(\vec{p}- \vec{p}\,')$
is the same as the one-body operators $a,a^\dagger$ (except for the normalization  factor $2p_0$). So the state
$|p\rangle$ which is a dressed state in terms of the bare operators $a^\dagger,b^\dagger$, can be interpreted, in this
sense,  as an elementary particle in terms of the operator $A^\dagger$.


\subsection {Perturbative case}\label{pert}
The perturbative case is simply obtained from the preceding one by omitting $\phi_3$. Taking into account that from
eq. (\ref{eq6}) we have $\Gamma_1/(m^2-p^2)=\phi_1$, we find:  $\Gamma_2=g\phi_1$. Substituting it into (\ref{psi2}),
we find $\phi_2$ and then, by eq. (\ref{nor2.b}),  we obtain $N_2$:
	\begin{equation}\label{I2}
	N_2=\phi_1^2 g^2 I_2, \quad \mbox{where }\quad I_2=\frac{1}{16\pi^3}\int \frac{\sqrt{s}}{(s_{12}-m^2)^2} \frac{d^3q}{\varepsilon(q,m)\varepsilon(q,\mu)}.
	\end{equation}
We thus find the ratio:
	\begin{equation}\label{rat}
	\frac{N_2}{N_1}=16\pi m^2\alpha I_2\approx 0.38\alpha
	\end{equation}
with $\alpha=\frac{g^2}{16 \pi m^2}$ . The numerical value of $I_2$ is given for $m=0.94$, $\mu=0.14$.  The integral
$I_2$ is logarithmically divergent for $\mu\to 0$.  The two-body contribution is rapidly decreasing, when the mass of the
intermediate particle increases.


\section{Renormalization of the coupling constant}\label{rcc}
As mentioned above, the mass renormalization counterterm $\delta m^2$, for the state vector incorporating $N$ and
$N\pi$ states only is given by eq.  (\ref{eq10}) and coincides with the perturbative result.  For the state vector
incorporating the states $N$, $N\pi$ and $N\pi\pi$, it is determined by eq.  (\ref{eq11a}).  This renormalization constant
is infinite when the cutoff tends to infinity.

The coupling constant is also renormalized, though its renormalization is finite for the particular scalar system we are
interested in the present study.  We  show below how this renormalization can be carried out.  This renormalization is
a ``by-product'' of the renormalization of the wave function fulfilled in the previous section.  Let us consider first the case
of the state vector containing $N$ and $N\pi$ states only.


\subsection{Truncation to $N$ and $N\pi$ states}
In this approximation, the normalization factor $N_2$ is given by eq. (\ref{I2}). We thus get:
	\begin{equation}\label{ren1}
	Z=N_1+N_2=\phi_1^2 Z_1\quad\mbox{with}\quad Z_1=1+g^2I_2,
	\end{equation}
where $I_2$ is given by eq. (\ref{I2}). After renormalization the wave function $\phi_2$ turns therefore into:
	\begin{equation}\label{ren2}
	\phi_2=\frac{g\phi_1}{s-m^2}\to \phi^{ren}_2=\frac{g}{\sqrt{Z_1}(s-m^2)}= \frac{g_{ren}}{s-m^2}
	\end{equation}
where we introduced the renormalized coupling constant
	\begin{equation}\label{ren3}
	g_{ren}=g/\sqrt{Z_1}.
	\end{equation}
This value  of $g_{ren}$ can be also  represented as the residue of the two-body wave function at $s=m^2$, i.e.
the value of $\Gamma_2(q,z)$ at the nonphysical value  of $q$ corresponding to  $s=m^2$.

One can alternatively define $g_{ren}$ from the $\pi N$ scattering amplitude determined by $N$ exchange in
the $s$-channel:
	\begin{equation}\label{ren4}
	F=\frac{g^2}{m^2-p^2}+\frac{g^2}{m^2-p^2}(\delta m^2 +g^2\Sigma(p^2))\frac{g^2}{m^2-p^2}+ \ldots =\frac{g^2}{m^2-p^2-\delta m^2-g^2\Sigma(p^2)}
	\end{equation}
where $p^2=(k_1+k_2)^2-2(\omega\cd p)\tau$.  Near the pole $p^2=m^2$ we get:
	\begin{equation}\label{ren5}
	F=\frac{g^2}{(m^2-p^2)(1+g^2\left.d\Sigma(p^2)/dp^2\right|_{p^2=m^2})} =\frac{g_{ren}^2}{m^2-p^2}
	\end{equation}
where we introduced $g_{ren}$ by eq. (\ref{ren3}), but with $Z_1$ given by:
	\begin{equation}\label{REN6}
	Z_1=1+g^2\left.\frac{d\Sigma(p^2)}{dp^2}\right|_{p^2=m^2}.
	\end{equation}
Taking eqs. (\ref{eq7a}) for $\Sigma$ one can easily check that the equations (\ref{ren1}) and (\ref{REN6})
determine the same $Z_1$.

Note that the renormalized coupling constant $g_{ren}$ is  finite and it is always smaller than the bare one $g$.
When the bare constant $g$ tends to infinity, the constant $g_{ren}$ remains however finite, but it  reaches its
maximal value. Expressed in terms of $\alpha=g^2/(16\pi m^2)$ it has the form:
	$$ \alpha_{ren}^{max}=1/(16\pi m^2 I_2). $$
It does not depend on $g$. According to (\ref{rat}), its numerical value for $m=0.94$ and $\mu=0.14$  is:
$\alpha_{ren}^{max}=1/0.38=2.63$.


\subsection{Truncation to $N$, $N\pi$ and $N\pi\pi$ states}
In this case the renormalized wave functions are given by eq. (\ref{nor1b}) with $Z$ determined by eqs.
(\ref{nor1a},\ref{nor2}). Note that the renormalization constant $Z$ can still be represented in the form (\ref{REN6})
with $\Sigma(p^2)$ determined by all the irreducible contributions (see appendix \ref{zall}).

Restricting ourselves to the system considered above and having found the renormalized state vector $|p\rangle$,
we can get all the physical information (we can for example calculate the electromagnetic form factors, if the particles
are charged).  So we do not need in practice to define and calculate the renormalized coupling constant $g_{ren}$.
However, it is useful to calculate it for further generalization to the case of particles with spin, when the charge
renormalization constant will become infinite.

As we already mentioned, the standard definition of the coupling constant is the residue of the wave function $\phi_2$
at $s=m^2$, i.e., the value of $\Gamma_2(q,z)$ at the non-physical value of $q=i\kappa$ corresponding to $s=m^2$.
For a  non-relativistic bound state  calculation for instance, it is given by  $\kappa=\sqrt{m|\epsilon_b|}$, where
$\epsilon_b$ is the binding   energy of the bound state.  In coordinate space it is the coefficient of the asymptotical
behavior of the wave function $\psi(r\to \infty)\propto \exp(-\kappa r)$.  Since we calculate $\Gamma_2(q,z)$ numerically
in the physical region for $q$, it is not easy to find numerically with enough accuracy its interpolation into the
non-physical region.  However, we do not need to choose the renormalization point $s=m^2$, but we can choose any
other renormalization point.  We can define for instance the renormalized coupling constant as the value of
$\Gamma_2(q,z)$ at $q=0$:
	\begin{equation}\label{ren7}
	\tilde{g}_{ren}^2=\Gamma^{ren}_2(q=0,z).
	\end{equation}
Note that  $\Gamma_2(q=0,z)$ does not depend on $z$.  In the case of $N+N\pi$ intermediate states $\Gamma_2$
does not depend on $q$ and the renormalized coupling constants $\tilde{g}_{ren}$ and $g_{ren}$ coincide with each
other.


\section{Numerical results}\label{num}
Both methods to solve eqs.(\ref{eq5_3_1}) are used to cross-check our results.  For regularization purposes, we
introduce a cutoff $L$, i.e. integrate in (\ref{eq11a}) over modulae of all the relative three-momenta $q$ until $q \leq L$.
Note that this cut-off procedure preserves rotational invariance.  For the masses, we choose the nucleon and pion
masses: $m=0.94$ GeV and $\mu=0.14$ GeV. The integration over the azimuthal angle is done analytically.
The equation is reduced to a matrix form by discretizing the integral.  The convergence of the integrals is already
obtained for $30$ points in the variable $q$ and $15$ points in the variable $z$.  The points in $q$ where not taken
equally spaced, but with a spacing proportional to $h^2$, where $h$ is the equal spacing in the variable $\sqrt{q}$.
In the first method, the eigenvalue $\lambda(\delta m^2)$ of the matrix is found numerically and $\delta m^2$ is fixed
to get $\lambda \equiv 1$ to less than $0.5\%$.  For the second method, $\delta m^2$ is calculated by a standard
iteration procedure, starting from the perturbative result.  Since both methods give identical results to less than $1\%$
we only quote the results obtained with the second method.

The results for  the dimensionless coupling constant $\alpha=g^2/(16\pi m^2)=3$ and for different values of the cutoff
parameter $L$ are  shown in table \ref{tab1}. We denote  by $\delta m^{2}/\delta m^{2}_0 $ the ratio of the value
$\delta m^{2}$ found from  eq. (\ref{eq11a}), to the perturbative value $\delta m^{2}_0 $, given by eq. (\ref{eq10}).
The values $N_{1,2,3}$ denote the contributions of the corresponding Fock sectors to the normalization of the state
vector. In order to characterize the wave function quantitatively, we calculate also the average value of the (absolute)
relative momentum  $<q>$, normalized to the two-body Fock component:
	\begin{equation}
	<q>=\frac{1}{N_{2}} \frac{1}{2(2\pi)^3}\int \phi_2^2(\vec{q},\vec{n}) \frac{[\varepsilon(q,m)+\varepsilon(q,\mu)]} {\varepsilon(q,m)\varepsilon(q,\mu)} \ q \ d^3q\ .
	\end{equation}

	\begin{table}[!htb]
	\begin{center}
	\begin{tabular}{||r||l|l|l|l|l|l||}
	\hline $L$      &  1  & 5  & 10   & 50 & 100  & 200\\
	\hline $\delta m^{2}_0$          &-1.234 &-3.67  &-4.82  &-7.54  &-8.71  &-9.88 \\
	\hline $\delta m^{2}/\delta m^{2}_0$          & 1.095 & 1.052 & 1.040 & 1.025 & 1.022 & 1.020 \\
	\hline $N_1$    & 0.377 & 0.329 & 0.327 & 0.326 & 0.326 & 0.326 \\
	\hline $N_2$    & 0.441 & 0.455 & 0.456 & 0.458 & 0.458 & 0.458 \\
	\hline $N_2/N_{1}$ & 1.183 & 1.380 & 1.395 & 1.407 & 1.405 & 1.405\\
	\hline $N_3$    & 0.182 & 0.216 & 0.217 & 0.216 & 0.216 & 0.216 \\
	\hline $<q>$    & 0.366 & 0.538 & 0.565 & 0.588 & 0.591 & 0.593\\
	\hline
	\end{tabular}
	\end{center}
	\caption{\label{tab1} Numerical results for $\alpha=3$, as a function  of the cut-off $L$.}
	\end{table}

One clearly sees that though $\delta m^{2}$ increases logarithmically, the contributions of the Fock components
$N_1,N_2,N_3$, as well as the average momentum $<q>$  in the two-body Fock component become stable after $L=5$.
This  means that we indeed found numerically the renormalized solution for the wave function.

One can see also that the nonperturbative value of $\delta m^{2}$  is very close to the perturbative one $\delta m^{2}_0$.
We have checked that the function $\Gamma_2(q,z)$ is almost constant, as expected for the perturbative solution. The
solution for $\delta m^2$ remains very close to the perturbative  result also for higher values of the coupling constant.
The results for $L=200$ and for different $\alpha$ between $1$ and $1000$ are shown in table \ref{tab2}.  One can
see that $\delta m^{2}$ is very close to the perturbative value even for extremely large coupling constant $\alpha=1000$.
As expected, the three-body sector dominates, when the coupling constant increases.  However, the ratio $N_2/N_1$ is
still close to its perturbative value $\left(N_2/N_1\right)_0$, given by eq.  (\ref{rat}).

	\begin{table}[!htb]
	\begin{center}
	\begin{tabular}{||r||l|l|l|l|l|l||}
	\hline $\alpha$  & 1     & 3    & 10    & 30   & 100 & 1000 \\
	\hline $\delta m^{2}/\delta m^{2}_0 $          & 1.011 &1.019  & 1.029  & 1.035 & 1.040 & 1.043\\
	\hline $N_1$    & 0.661 &0.325  & 0.081  & 0.016 & 2. E-3 & 3. E-5 \\
	\hline $N_2$    & 0.292 & 0.457 & 0.366  & 0.186  & 0.068 & 8. E-3 \\
	\hline $N_3$    & 0.047 & 0.218 & 0.553  & 0.798 & 0.930 & 0.992 \\
	\hline $N_2/N_1$ & 0.441  & 1.41 & 4.51   & 11.91 & 32.9  & 230 \\
	\hline $\left(N_2/N_1\right)_0$           & 0.38   & 1.14 & 3.8    & 11.4  & 38    & 380  \\
	\hline
	\end{tabular}
	\end{center}
	\caption{\label{tab2} Same as Table 1, but with $L$ fixed to $200$,  and $\alpha$ is varied.}
	\end{table}

The reason that the solution is close to the perturbative one lies in the super-renormalizability of the scalar theory.
The third term in eq.  (\ref{eq11a}) converges and does not require any cutoff, whereas the first two terms are divergent
and dominate.  Therefore, a very small departure of $\delta m^2$ from its perturbative value $\delta m_{0}^2$ can
accommodate the finite higher order correction.  If the third term can be neglected, the equation is approximately
satisfied with the perturbative value of $\delta m^{2}$.  In order to show that it is indeed so, we introduce the cutoff in
the two divergent terms, but do not introduce it in the third, convergent one.  The results for $\alpha=3$ are shown in
table \ref{tab3}.  For small enough value of $L$ the two first terms are suppressed, and the solution drastically differs
from the perturbative on.  When $L$ increases, the solution becomes closer and closer to the perturbative one.

	\begin{table}[!htb]
	\begin{center}
	\begin{tabular}{||r||l|l|l|l||}
	\hline $L$      & 0.1  & 1    & 10  & 100  \\
	\hline $\delta m^{2}/\delta m^{2}_0$           &36.3 &1.55  &1.07 &1.03 \\
	\hline \end{tabular}
	\end{center}
	\caption{\label{tab3} Test calculation in which the cutoff $L$ is introduced only in the two first terms in
	eq. \protect{\ref{eq11a}} (see text), for $\alpha=3$.}
	\end{table}

We emphasize that the small deviation of the solution from the perturbative one, even for very large values of the
coupling constant, is just a property of the non-perturbative equation (\ref{eq11a}), which includes the contributions
(with one-, two- and three-body intermediate states) to all orders of $g$. This result cannot be justified in any
perturbative expansion in  terms of $g$.


\section{Conclusion}\label{concl}
In a first attempt to address the question of non-perturbative renormalization in CLFD, we have investigated in this
study a simple, but nevertheless meaningful, model based on two scalar particles. This model is reminiscent of the
structure of the physical nucleon in the low energy regime, in terms of bare nucleons coupled to pions.

Using the nice features of CLFD, we have first derived the general eigenstate equation that should be used in order
to calculate any physical state vector.  We emphasize here that this equation is quite general and is not restricted
to the case of scalar particles nor to the restricted Fock space we consider in this study.  It should therefore also be
used when solving more complex systems as QED or QCD.

The results we obtained for the simple scalar model, with a restricted Fock space expansion up to three particles, are
quite encouraging. We obtained, numerically, a renormalized solution using a simple mass counter term. Surprisingly
enough, this counter term is not very different from the perturbative, logarithmically divergent, mass counterterm, even
for very large values of the coupling constant. We traced back this feature to the nature of our scalar model we start
from,  which is super-renormalizable. This result however does not imply that higher Fock states are negligible. We
find that the first non-trivial Fock component gets larger and larger as the coupling constant increases.

The direct generalization of this study is the investigation of non-perturbative renormalization in scalar QED, following
the study of perturbative renormalization in QED already done in ref. \cite{kar00} in CLFD. This will be the subject of
a forthcoming publication.

\section*{Acknowledgement}
One of the authors (V.A.K.) is sincerely grateful for the warm hospitality  of Laboratoire de Physique Corpusculaire,
Universit\'e Blaise Pascal, in Clermont-Ferrand, where this work was performed. This work is partially supported by
the grant No. 99-02-17263 of the Russian Fund for Basic Researches, and by the Russian-French "PICS" research
contract No. 1172.


\appendix

\section{Kinematics}  We show in this appendix, how to express the propagator (\ref{eq12}) in terms of the variables
$\vec{q},\vec{q}\,',\vec{n}$.

The two-body wave function depends on the following four-vectors:
	\begin{equation}\label{ap2}
	\phi=\phi(k_{1}, k_{2}, p,\omega\tau), \quad k_{1}+  k_{2}=  p+\omega\tau\ .
	\end{equation}
We introduce the variables:
	\begin{equation}\label{sc4}
	\vec{q}= L^{-1}({\cal P})\vec{k}_1 = \vec{k}_1 - \frac{\vec{\cal P}}{\sqrt{{\cal P}^2}}[k_{10} - \frac{\vec{k}_1\cd\vec{{\cal P}}}{\sqrt{{\cal P}^2}+{\cal P}_0}]\ ,
	\end{equation}
	\begin{equation}\label{sc5}
	\vec{n} = L^{-1}({\cal P})\vec{\omega}/|L^{-1}({\cal P}) \vec{\omega}| = \sqrt{{\cal P}^2} L^{-1}({\cal P}) \vec{\omega}/\omega\cd p\ ,
	\end{equation}
where
	\begin{equation}\label{calp}
	{\cal P} = p + \omega\tau\ ,
	\end{equation}
and $L^{-1}({\cal P})$ is the Lorentz boost. The wave function under the integral (see the last diagram in fig. 3) depends
 on $k'_{1}, k'_{2}, p,\omega\tau'$, and, correspondingly, on the variable:
	\begin{equation}\label{sc4a}
	\vec{q}\,'= L^{-1}({\cal P}')\vec{k}\,'_1 \ ,
	\end{equation}
where
	\begin{equation}\label{calpa}
	{\cal P}' = p + \omega\tau'.
	\end{equation}
In order to obtain $\Pi(\vec{q},\vec{q}\,',\vec{n},p^{2})$, we should express the four-momenta
$k_{1}',k_{2},\omega\tau'$ in $(k_{1}'-k_{2}-\omega\tau')^{2}$ in terms of $\vec{q},\vec{q}\,',\vec{n}$.  Since under
the Lorentz transformations and the rotations of the four-vectors the variables $\vec{q},\vec{q}\,',\vec{n}$ are rotated
only, the expression for the scalar $\Pi(\vec{q},\vec{q}\,',\vec{n},p^{2})$ does not depend on the system of reference,
and, hence, can be found in the most convenient one.  We find it in the system where
	$$\vec{{\cal P}}' =\vec{k}\,'_{1}+\vec{k}\,'_{2}=\vec{p} + \vec{\omega}\tau'=0\ .$$

In this system $\vec{k}\,'_{1}=\vec{q}\,'$ and $k'_{10}=\varepsilon(\vec{q}\,',m)$. From the conservation law the
vector $\vec{\cal P}$ is expressed as: $\vec{\cal P}=\vec{k}_{1}+\vec{k}_{2}=\vec{\omega}(\tau-\tau')$. We
have:
	\begin{equation}\label{ap7}
	\tau=\frac{s-p^{2}}{2(\omega\cd p)},\quad \tau'=\frac{s'-p^{2}}{2(\omega\cd p)}\ ,
	\end{equation}
where $\sqrt{s}=\varepsilon(\vec{q},m)+\varepsilon(\vec{q},\mu)$,
$\sqrt{s'}=\varepsilon(\vec{q}\,',m)+\varepsilon(\vec{q}\,',\mu)$. Since $\vec{\omega}=\omega_{0}\vec{n}$ and
$\omega\cd p=\omega\cd(k'_{1}+k'_{2})=\omega_{0} (\varepsilon(\vec{q}\,',m) +\varepsilon(\vec{q}\,',\mu))=\omega_{0}\sqrt{s'}$,
we find
	\begin{equation}\label{ap3}
	\vec{\cal P}=\vec{n}\frac{s-s'}{2\sqrt{s'}},\quad {\cal P}_{0}=\frac{s+s'}{2\sqrt{s'}}\ .
	\end{equation}
One can check that ${\cal P}^{2}=s$. According to eqs. (\ref{sc4}) and (\ref{sc4a}), the variables $\vec{q}$ and
$\vec{q}\,'$ are defined by different Lorentz boosts. From (\ref{ap3}) one can see that this difference is the boost in
the direction of $\vec{n}$.  This boost does not change the unit vector $\vec{n}$.

Now let us find $\vec{k}_{2}$. It is obtained by reverting (\ref{sc4}):
	\begin{equation}\label{ap4}
	\vec{k}_{2}= L({\cal P})(\vec{-q}) = -\vec{q} + \frac{\vec{\cal P}}{\sqrt{{\cal P}^2}}[\varepsilon(\vec{q},\mu) - \frac{\vec{q}\cd\vec{{\cal P}}}{\sqrt{{\cal P}^2}+{\cal P}_0}] \ .
	\end{equation}
Eq. (\ref{ap4}) is obtained from (\ref{sc4}) by replacing in the r.h.s $\vec{k}_{1}$ by $-\vec{q}$ and by changing the sign
at $\vec{\cal P}$.

Substituting here eqs. (\ref{ap3}) for ${\cal P}$, we find:
	\begin{equation}\label{ap5}
	\vec{k}_{2}= -\vec{q}+ \vec{n}\frac{s-s'}{2\sqrt{ss'}}\left( \varepsilon(\vec{q},\mu) -\vec{n}\cd\vec{q}\frac{\sqrt{s}-\sqrt{s'}} {\sqrt{s}+\sqrt{s'}}\right)\ .
	\end{equation}
and similarly for $k_{20}$:
	\begin{equation}\label{ap6}
	k_{20}=\frac{\varepsilon(\vec{q},\mu){\cal P}_{0}}{\sqrt{{\cal P}^2}}- \frac{\vec{q}\cd\vec{\cal P}}{\sqrt{{\cal P}^2}}= \varepsilon(\vec{q},\mu)\frac{s+s'}{2\sqrt{ss'}}- \vec{n}\cd\vec{q} \frac{s-s'}{2\sqrt{ss'}}\ .
	\end{equation}
From (\ref{ap7}) we get:
	$$\vec{\omega}\tau'=\vec{n}\frac{s'-p^{2}}{2\sqrt{s'}},\quad \omega_{0}\tau'=\frac{s'-p^{2}}{2\sqrt{s'}}\ . $$
Substituting the above expressions for the four-momenta into $(k_{1}'-k_{2}-\omega\tau')^{2}$, we find that the
four-vector squared in the denominator of the propagator (\ref{eq12}) is expressed in terms of the variables
$\vec{q},\vec{q}\,',\vec{n}$ as follows:
	\begin{eqnarray}\label{ap1}
	(k_{1}'-k_{2}-\omega\tau')^{2}
	&=&\left[\varepsilon(\vec{q}\,',m) -\varepsilon(\vec{q},\mu)\frac{s+s'}{2\sqrt{s s'}} +\vec{n}\cd \vec{q} \frac{s-s'}{2\sqrt{s s'}}-\frac{s'-p^{2}}{2\sqrt{s'}}\right]^{2} \nonumber\\
	&-&\left[\vec{q}\,'+\vec{q}-\vec{n}\frac{s-s'}{2\sqrt{s s'}} \left(\varepsilon(\vec{q},\mu)-\vec{n}\cd\vec{q}\frac{\sqrt{s}-\sqrt{s'}} {\sqrt{s}+\sqrt{s'}}\right) -\vec{n}\frac{s'-p^{2}}{2\sqrt{s'}}\right]^{2}.
	\end{eqnarray}


\section{Calculation of $\phi_1$ and $\phi_3$}\label{psi13}
The Fock components $\phi_{1}$ and $\phi_{3}$ are calculated as follows. The wave function $\phi_2$ in
(\ref{nor2.b}) is related to the vertex functions $\Gamma_2$  by eq. (\ref{psi2}). Having found numerically the
vertex function $\Gamma_2(\vec {q},\vec{n})=\Gamma_2(q,z)$ from the equation (\ref{eq11a}), we then find by
(\ref{psi2}) the Fock component $\phi_2$.

With eqs. (\ref{eq5_3.a}), reduced to the first equation in (\ref{eq6}), we express $\Gamma_1$ through $\Gamma_2$
in the limit $p^2\to m^2$,  and then find the component $\phi_1$:
	\begin{equation}\label{psi1}
	\phi_1=-\frac{g}{\delta m^2} \int\Sigma_i(\vec{q}\,',p^2) \Gamma_{2}(\vec{q}\,',\vec{n})\frac{d^{3}q'}{(2\pi)^{3}}\ .
	\end{equation}
With eqs. (\ref{eq5_3.c}), we express $\Gamma_3$ through $\Gamma_2$ and then find $\phi_3$:
	\begin{equation}\label{psi3}
	\phi_3(\vec{q}_1,\vec{q}_2,\vec{q}_3,\vec{n})= \frac{g\phi_2(q',z')}{(s_{123}-m^2)(1-x_2)} +\frac{g\phi_2(q'',z'')}{(s_{123}-m^2)(1-x_3)}\ .
	\end{equation}
Here $\phi_2(q',z')$ is $\phi_2(k_1',k_2,p,\omega\tau_2')$  represented through the relative momentum, and
$\phi_2(q'',z'')$ is  $\phi_2(k_1'',k_2,p,\omega\tau_2'')$.

The wave function $\phi_2$  in the r.h.s. of eq. (\ref{psi3}) depends on the variables defined in the center of mass of
the two-body subsystem.  We shall now express these variables in terms of the three-body relative momenta
$\vec{q}_1,\vec{q}_2,\vec{q}_3$.

The variable $q'$ is related to $s_{12}'$ by $s_{12}'=\left[\varepsilon(q',m)+\varepsilon(q',\mu)\right]^2$. We  thus get:
	$$ {q'}^2=\left[m^4+(\mu^2-s_{12}')^2-2m^2(\mu^2+s_{12}')\right]/(4s_{12}')\ . $$
where, by taking the first of eqs. (\ref{conserv}) squared:
	$$ s_{12}'=s_{123}-\frac{s_{13}-m^2}{1-x_2}\ , $$
with:
	\begin{eqnarray}\label{s123}
	s_{123}&=&(\varepsilon(q_1,m)+\varepsilon(q_2,\mu)+\varepsilon(q_3,\mu))^2\ , \nonumber\\
	s_{13}&=&(k_1+k_3)^2=(\varepsilon(q_1,m)+\varepsilon(q_3,\mu))^2-q_2^2\ , \nonumber\\
	x_{1}&=&\frac{\varepsilon(q_1,m)-\vec{n}\cd \vec{q}_1}{\sqrt{s_{123}}},\quad
	x_{2,3}=\frac{\varepsilon(q_{2,3},\mu)-\vec{n}\cd  \vec{q}_{2,3}}{\sqrt{s_{123}}}\ .
	\end{eqnarray}
We use the fact that $\vec{q}_1+\vec{q}_2+\vec{q}_3=\vec{0}$.

The value of $z'$ is found by comparing expressions (\ref{s123}) for $x_1$ with:
	$$ x_1=\frac{\varepsilon(q',m)-z' q'}{\varepsilon(q',m)+\varepsilon(q',\mu)}\ . $$
This gives:
	$$ z'=\left[\varepsilon(q',m)-x_1\left[\varepsilon(q',m)+\varepsilon(q',\  mu)\right] \right]/q'\ , $$
where $x_1$  is given by (\ref{s123}) in terms of $\vec{q}_1,\vec{q}_2,\vec{q}_3$.

The values of $q'',z''$ are found from the equations:
	$$ {q''}^2=\left[m^4+(\mu^2-s_{12}'')^2-2m^2(\mu^2+s_{12}'')\right]/(4s_{12}'')\ ,$$
	$$ z''=\left[\varepsilon(q'',m)-x_1(\varepsilon(q'',m) +\varepsilon(q'',\mu))\right]/q''\ , $$
where:
	$$ s_{12}''=s_{123}-\frac{s_{12}-m^2}{1-x_3}\ , $$
with  $s_{12}=(k_1+k_2)^2=\left[\varepsilon(q_1,m)+\varepsilon(q_2,\mu)\right]^2-q_3^2$.


\section{Proof of eq. (\ref{REN6}) in the general case}\label{zall}
We  show below that the renormalization constant $Z$ can  still be represented in the form (\ref{REN6}) with
$\Sigma(p^2)$ determined by all the irreducible contributions. In this case, the number of irreducible diagrams containing
two- and three-body intermediate states is infinite. Some of them are shown in fig. \ref{fig2}. Consider, for example, a
contribution containing $n$ successive intermediate states. The corresponding amplitude contains the factor $1/\tau_i$
for any of  these states. Like in eq. (\ref{eq7a}), any factor  $1/\tau_i$ turns into $1/(s_i-p^2)$, where $s_i$ is the invariant
energy of a given  intermediate state and $p$ is the external incoming/outgoing momentum. So, indicating only these
factors, we represent this contribution to   $\Sigma(p^2)$ as:
	$$ \Sigma_n(p^2)=\int \prod_{i=1,...,n}\frac{1}{s_i-p^2}\ldots $$
where the dots include all the integrations with the corresponding measures. Derivative over $p^2$ gives the factor
$1/(s_i-p^2)^2$.

Now consider the graphs which are the same at the right to a given $i$-th intermediate state and differ from each
other by the contributions at the left to this state. The infinite sum of them determines the amplitude of the virtual
transition from the initial state $N$ to the states $N\pi$ or $N\pi\pi$. It is the vertex $\Gamma_2$ (if the state $i$ is
the two-body state), or $\Gamma_3$ (if the state $i$ is the three-body state). We can then take the sum over all the
contributions to the right of this given state $i$ and again obtain $\Gamma_2$ or $\Gamma_3$. So, the result has the
form
	$$ \frac{d\Sigma(p^2)}{d p^2}= \int \frac{\Gamma_2^2}{(s_i-p^2)^2}\ldots +\int \frac{\Gamma_3^2}{(s_i-p^2)^2}\ldots =\int \phi_2^2 \ldots +\int \phi_3^2\ldots =N_2+N_3 $$
and, after extraction of the common factor $N_1$, we recover eq. (\ref{REN6}).


\end{document}